\pdfoutput=1

\documentclass[final,1p,times]{elsarticle} 
\usepackage{graphicx} 
\usepackage{amssymb} 
\usepackage{amsthm} 
\usepackage{lineno} 
\usepackage{subfigure}

\journal{Nuclear Physics A} 
\begin{document} 

\begin{frontmatter} 


\title{Heavy Flavor Production and Energy Loss with Two-Particle Correlations at PHENIX}

\author{Tatia Engelmore for the PHENIX Collaboration}

\address{Columbia University, New York, New York 10027, USA and Nevis Laboratories, Irvington, New York 10533, USA }

\begin{abstract} 
Heavy quarks are a valuable probe of the hot, dense medium created in a heavy ion collision, and are an important test of proposed mechanisms of energy loss.  It was discovered that single non-photonic electrons are suppressed at a similar level to light hadrons, implying a comparable level of energy loss between light and heavy partons.  Because theory has had a difficult time explaining the level of heavy quark energy loss, it is crucial to better understand charm and bottom suppression.  Electron-hadron correlations have been used at PHENIX to study heavy flavor in both $p$+$p$ and Au+Au collisions.  In $p$+$p$ the ratio of charm to bottom production has been measured using mass correlations through a partial reconstruction of the D meson.  Electron-hadron angular correlations have also been used to study medium modification of heavy flavor, and we see hints of energy loss effects.  A complementary study of  correlated electron-muon pairs provides a clean measurement of heavy flavor production in a rapidity range not yet studied.  
\end{abstract} 

\end{frontmatter} 



\section{Introduction}\label{intro}

Two-particle correlation studies at RHIC have yielded valuable information on the interaction of jets produced in nucleus-nucleus collisions with the hot nuclear matter.  These correlations, when compared against the baseline from proton-proton collisions, show powerful evidence of the modification of jet structure within the medium \cite{jets}.  Until now, jet interactions at PHENIX have been studied using correlations with light hadrons; a natural next step, then, is to understand the medium modification of heavy-flavor jets.  The suppression of single leptons from heavy flavor decay has been shown to be similar to the suppression of $\pi^{0}$ \cite{single_e}, meaning that heavy quark energy loss in the medium is significant.  Therefore it is of interest to determine the effects of the medium on the structure of heavy flavor jets.  

PHENIX measures charm and bottom mesons mainly through the electron and muon decay products.  In order to isolate double heavy flavor decays we have used correlations of electrons and hadrons coming from the same $c\bar{c}$ or $b\bar{b}$ pair. The heavy flavor electrons are isolated through a statistical subtraction of the background.  Mass correlations are used to give us a measure of the ratio of charm to bottom produced in the collision, while the correlation of $e$-$h$ in azimuthal angle allows us to study jet properties of heavy flavor.   Results from the PHENIX $e$-$h$ correlations in $p$+$p$ and in Au+Au are now presented.  A complementary analysis involves electron-muon correlations where both leptons are charm or bottom meson decay products.  Although the statistics are low, there is a clean heavy flavor signal.  Results are shown from the analysis in $p$+$p$ collisions.   Electron-muon correlations in $d$+Au will also be interesting in order to probe the modification of the gluon structure function in heavy ions.

\section{Electron-Hadron Correlations}

Because the majority of electrons measured at PHENIX are due to photonic sources, the heavy flavor signal must be isolated through a statistical subtraction.  What we measure is the inclusive yield of hadrons per trigger electron, $Y_{e_{inc}-h}$, which is proportional to both the heavy flavor and the photonic per-trigger yield:

\begin{equation}\label{inclusive_Y}
Y_{e_{inc}-h} = \frac{N_{e_{HF}}Y_{e_{HF}-h} + N_{e_{phot}}Y_{e_{phot}-h}}{N_{e_{HF}} + N_{e_{phot}}}
\end{equation}

\parindent0ex
From this we find the heavy flavor yield:

\begin{equation}\label{HF_Y}
Y_{e_{HF}-h} = \frac{(R_{HF} + 1)Y_{e_{inc}-h} - Y_{e_{phot}-h}}{R_{HF}}
\end{equation}
 
where $R_{HF} = \frac{N_{e_{HF}}}{N_{e_{phot}}}$ which was found in \cite{e_frac}.  

\parindent03ex 
The important quantity to determine is $Y_{e_{phot}-h}$.  This yield is primarily due to photonic electrons arising from the decay, $\pi^{0} \rightarrow \gamma\gamma$.  These decay photons then convert to electrons in the detector elements.  Two methods are used to determine the photonic electron-hadron yield.  In the first, the measured PHENIX inclusive photon spectrum is used as the input to a GEANT simulation; these photons are predominantly from Dalitz decays.  The electrons resulting from conversions are then reconstructed, and the mapping between the photon $p_{T}$ and the electron $p_{T}$ is determined.  For the second method, the measured $\pi^{0}$ spectrum is input into a Monte Carlo simulation, and the correspondence between the decay $\gamma$ $p_{T}$ and the $p_{T}$ of the electron resulting from the photon's conversion is found.  The relationships between the photons and their resultant electrons are used to determine weights in order to construct the $Y_{e_{phot}-h}$ from the $Y_{\gamma_{inc}-h}$:
\begin{equation}\label{weights}
Y_{e_{phot}-h}(p_{T, i}) = \sum_{j}w_{i}(p_{T, j})Y_{\gamma_{inc-h}}(p_{T, j})
\end{equation}

\begin{figure}[h]
\centering 
\subfigure[]{
\includegraphics[scale=0.45]{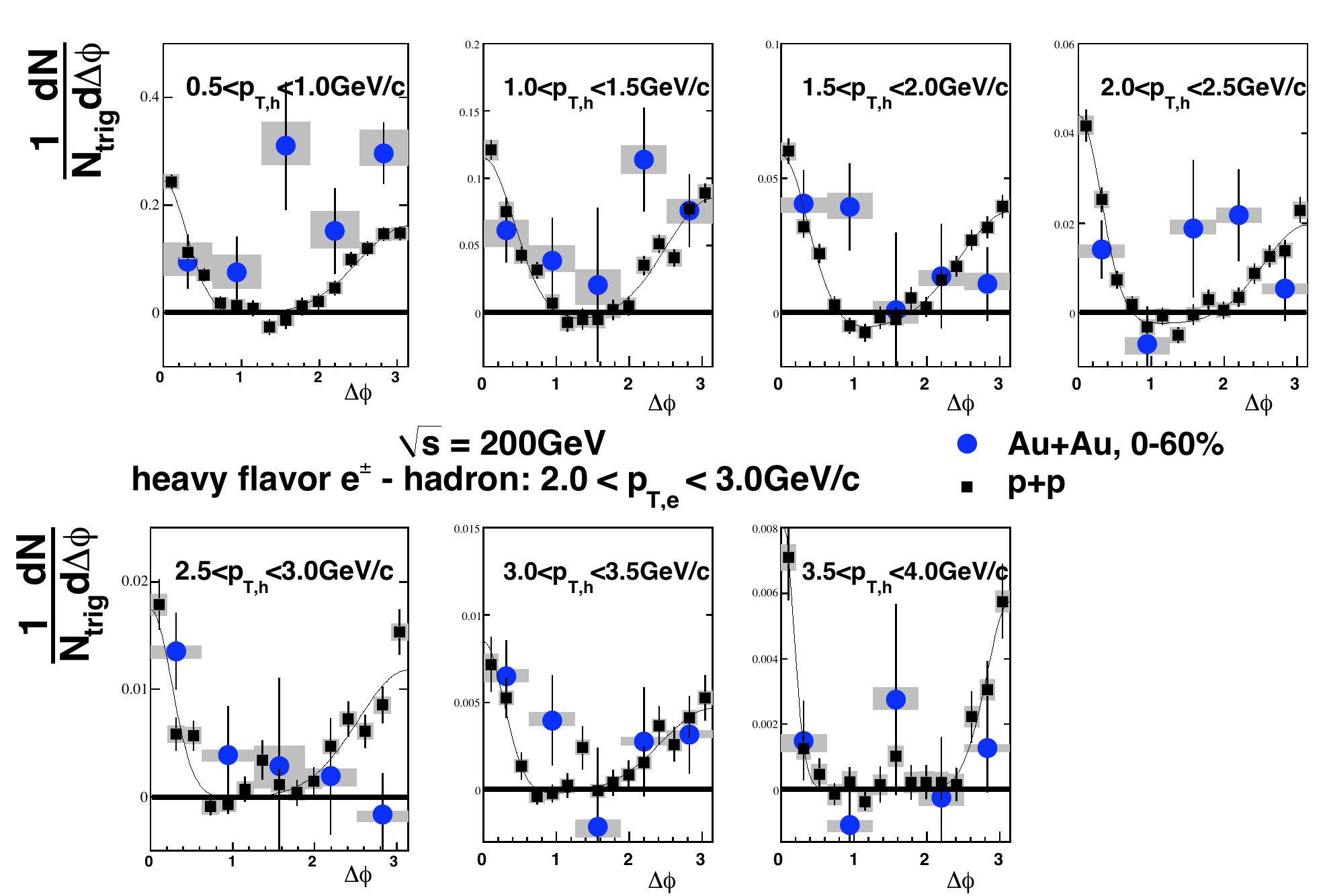}
\label{fig:hf_run7}
}
\subfigure[]{
\includegraphics[scale=0.25]{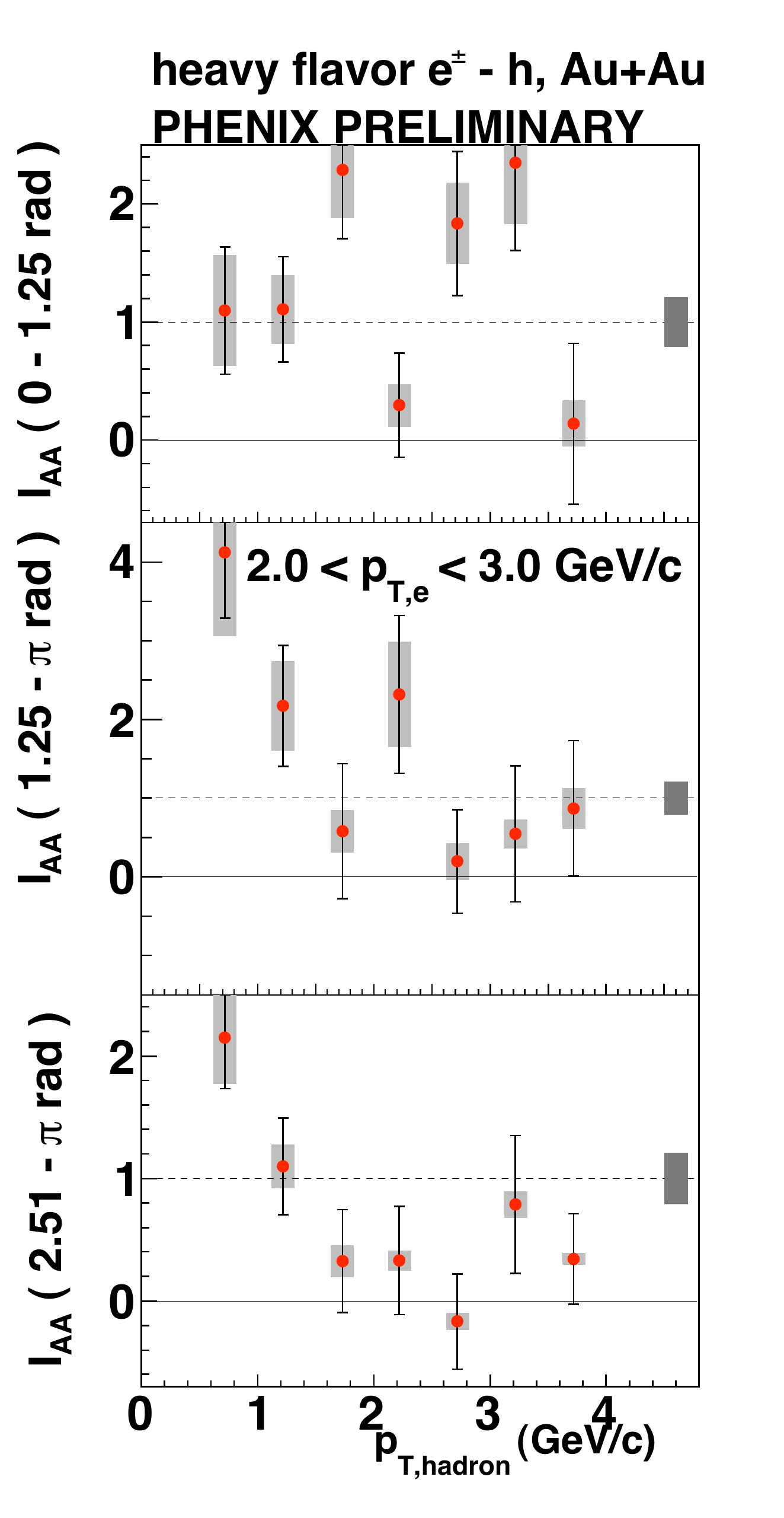}
\label{fig:iaa}
}
\caption[]{Results for $e$-$h$ correlations in Au+Au.  At left, angular correlations in both $p$+$p$ (black squares) and Au+Au (blue circles).  At right, $I_{AA}$ (ratio of per-trigger yield in $p$+$p$ to that in Au+Au) for $2.0 < p_{T, e} < 3.0$ GeV/c as a function of hadron $p_{T}$.  Top row corresponds to near side, middle to combined head and shoulder region, and bottom to head region alone.}
\label{fig:eh_2panel}
\end{figure}

The $e_{inc}-h$ and $\gamma-h$ correlations were then measured, and converted through the subtraction in Eq. \ref{HF_Y} to the heavy flavor per-trigger yield.  The angular correlations were then corrected for the restricted PHENIX acceptance using mixed events, and the combinatoric background was removed using a ZYAM subtraction \cite{ZYAM}.  Fig.~\ref{fig:hf_run7} shows the correlation functions for a trigger of $2.0 < p_{T} <3.0$ GeV/c with various associated $p_{T}$ bins.  Also, Fig.~\ref{fig:iaa} shows $I_{AA}$, which is the ratio of the per-trigger $e-h$ pair yield in Au+Au to that in $p+p$.  Results are shown for three regions in $\Delta\phi$, corresponding to (from top) the near side, the head and shoulder combined, and the head alone (for definitions see \cite{jet_shapes}).  A larger $I_{AA}$ for the head and shoulder region combined, rather than the head alone, gives a hint of a shoulder appearing in heavy flavor correlations.  Larger statistics are needed, though, to make this result conclusive.

While $e$-$h$ angular correlations measure inclusive heavy flavor, mass correlations have been used to separate charm from bottom.  This was found by partially reconstructing D mesons, and the ratio of charm to bottom as a function of $p_{T}$ was obtained \cite{c_to_b}.

\section{Electron-Muon Correlations}

Electron-muon correlations have long been proposed as a clean signal of charm pair production in heavy ion collisions \cite{charm_at_rhic}.  Because the heavy flavor signal consists of opposite-sign $e$-$\mu$ pairs rather than dielectrons or dimuons, there are no backgrounds from other physics processes such as Drell-Yan, thermal production, or resonance decays.  We are sensitive to the production of back-to-back charm pairs, mainly via gluon fusion \cite{vitev}. Since the signal is solely opposite-sign pairs, the majority of the background from combinatorics, light meson decays and photonic electrons is removed with a like-sign subtraction.

\begin{figure}[h]
\centering
\includegraphics[scale=0.4]{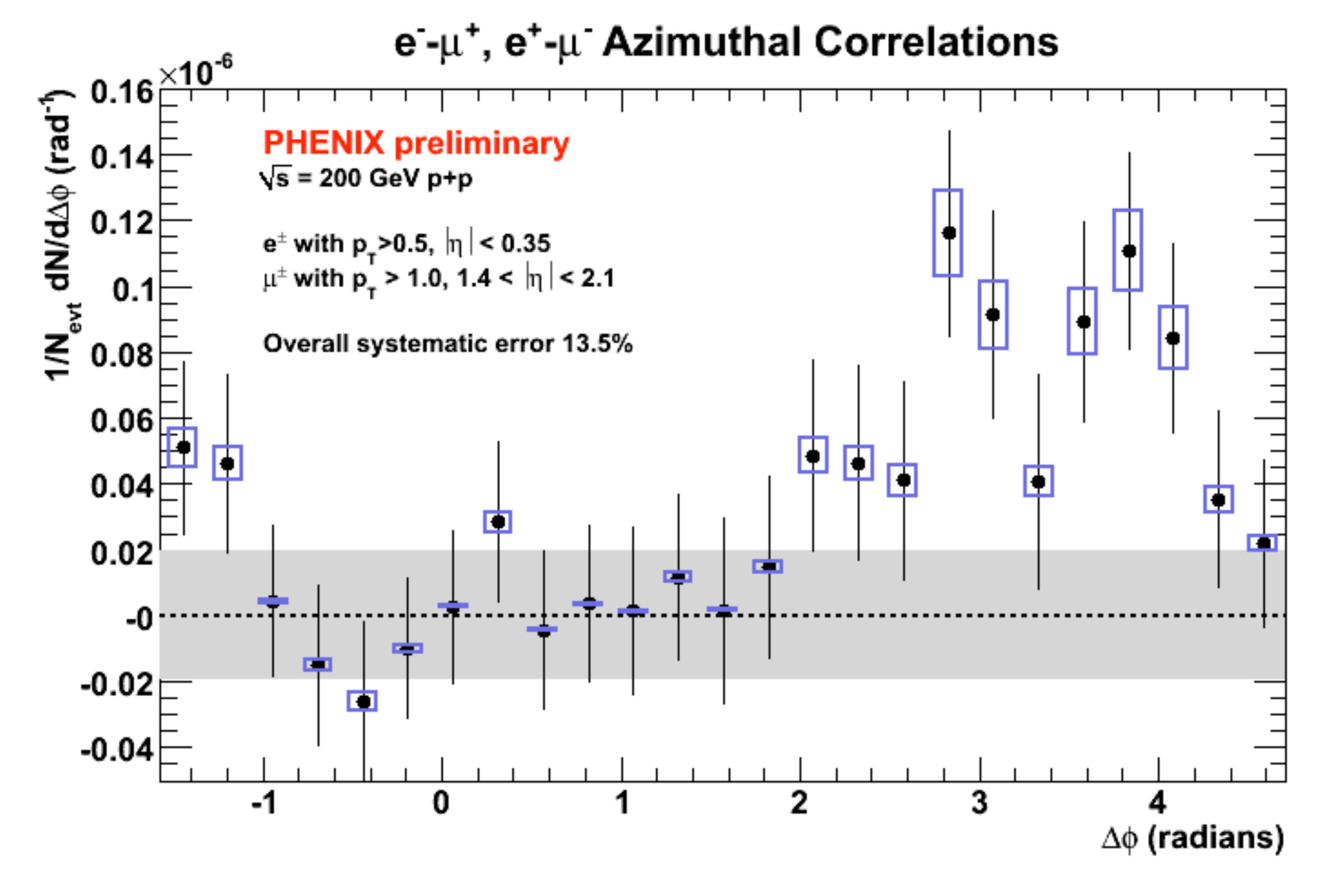}
\caption[]{Electron-muon correlations from heavy flavor in $p$+$p$.  Grey band corresponds to error from decay muon subtraction, blue boxes to error from ``punch-through" hadron subtraction. }
\label{fig:emu}
\end{figure}

The remaining backgrounds that need to be accounted for are due to standard backgrounds in the muon detectors \cite{single_mu}.  These include hadrons that survive the front muon absorber and appear as muons, as well as muons that arise from light meson decay rather than from heavy flavor.  While neither of these contribute significantly in the signal region ($\Delta\phi = \pi$), they affect the level of flat background to the angular correlation and must be subtracted.  Correlations of electrons with ``punch-through" hadrons were measured using tracks that stop within the PHENIX muon identifier.  For these tracks a clear muon stopping peak in the longitudinal momentum spectrum was seen.  Hadron tracks are those that fall outside of this peak.  To separate the correlations involving decay muons, a comparison of muon tracks with an event vertex near to and far from the detector was made.  This is because if light hadrons are created far from the detector, they have more of a chance to decay.  The contributions from both of these backgrounds were subtracted from the inclusive signal, and the result is shown in Fig. ~\ref{fig:emu}.

\section{Conclusions and Outlook}

The study of heavy flavor jet physics is important for understanding how heavy quarks propagate through the medium created in a heavy ion collision.  Heavy quarks have been demonstrated to lose similar amounts of energy as light mesons and are also shown to have elliptic flow \cite{v2}, so the modification of their jet structure is a crucial piece of information.  It will be interesting to compare the shape of these correlations to the structures found in hadron-hadron correlations, for example the shoulder and the ridge.  

The electron-hadron analysis requires a large background subtraction, so it was difficult to extract a clear measure of the signal.  With the addition of the silicon vertex upgrades to PHENIX, though, it will be easier to determine the heavy flavor signal and this measurement will become much cleaner.  Electron-muon correlations are limited by statistics, though this signal will be easier to measure with increased luminosity.  We have shown here a measurement of the $p$+$p$ baseline, which is necessary for future studies of these correlations in heavy ion measurements.  A study of the cold-matter effects on $e$-$\mu$ correlations in $d$+Au is underway.




\begin{thebibliography}{99} 
   
\bibitem{jets} A. Adare, et. al., \textit{Phys. Rev. C} 78, 014901 (2008)
\bibitem{single_e} A. Adare, et. al., \textit{Phys. Rev. Lett} 98, 172301 (2007)
\bibitem{e_frac} A. Adare, et. al., \textit{Phys. Rev. Lett} 97, 252002 (2006)
\bibitem{ZYAM} J. Jia, \textit{Nuclear Phys. A}, 783, 501c-506c (2007)
\bibitem{jet_shapes} A. Adare, et. al., \textit{Phys. Rev. C} 77, 011901 (2008)
\bibitem{c_to_b} A. Adare, et. al., [arXiv:0903.4851]
\bibitem{charm_at_rhic} S. Gavin, et. al., \textit{Phys. Rev. C} 54, 2606 (1996)
\bibitem{vitev} I. Vitev, et. al., \textit{Phys. Rev. D} 74, 054010 (2006)
\bibitem{single_mu} S. Adler, et al., \textit{Phys. Rev. D} 76, 092002 (2007)
\bibitem{v2} A. Adare, et. al, \textit{Phys. Rev. Lett} 98, 172301 (2007)
\end{thebibliography}
\end{document}